\documentstyle[tighten,floats,aps,epsf]{revtex}

\begin{document}
\title
{Deformations in $N=14$ isotones}

\author{Y. Kanada-En'yo}

\address{Institute of Particle and Nuclear Studies, \\
High Energy Accelerator Research Organization,\\
Ibaraki 305-0801, Japan}

\maketitle
\begin{abstract}
Systematic analysis of deformations in neutron-rich 
$N=14$ isotones was done based on the method of antisymmetrized 
molecular dynamics. 
The property of the shape coexistence in $^{28}$Si, which is known to have 
the oblate ground state and the prolate excited states, was successfully 
described.
The results suggest that 
the shape coexistence may occur also in neutron-rich $N=14$
nuclei as well as $^{28}$Si.
It was found that the oblate neutron shapes are favored because of 
the spin-orbit force in most of $N=14$ isotones.  
$Q$ moments and $E2$ transition strengths in the neutron-rich nuclei were 
discussed in relation to the intrinsic deformations, and
a possible difference between the proton and neutron deformations
in $^{24}$Ne was proposed.
\end{abstract}

\noindent

\section{Introduction}

The shape coexistence in $^{28}$Si has been studied for a long time.
At the early stage, the oblate and prolate solutions of 
Hartree-Fock calculations are associated with the ground $0^+$ state
and the $0^+_3$ state at 6.691 MeV, respectively\cite{Gupta67}.
Experimentally, the ground band is known to be oblate from the 
sign of the quadrupole moment of the $2^+_1$ state, 
while the the band-head $0^+_3$ in the 
prolate band has been identified by gamma transition 
measurements\cite{GLATZ}.

The coexistence of oblate and prolate states in $^{28}$Si
can be described by the shell gap at $N=14$ and $Z=14$ of 
Nilsson orbits in the deformed system \cite{Bohr75}. 
A variety of shapes is considered to 
arise from the nature of the shell structure in 
$N=14$ and/or $Z=14$ systems, where a half of the $sd$-shell is occupied.
It is natural to expect that shape coexistence phenomena may appear 
also in neutron-rich nuclei with $N=14$. 
From the coexistence of oblate and prolate states in $^{28}$Si, 
one may expect that this trend of the oblate and prolate neutron structure 
might survive in other $N=14$ isotones.
The primary interest in the present paper is on the deformations of 
the proton and neutron densities in neutron excess $N=14$ nuclei.
How does the oblate neutron structure in the ground band of $^{28}$Si
change with the decrease of proton number toward the neutron-rich region ? 
Whether or not the coexistence of oblate and prolate deformations 
occurs ? 
The neutron deformations should be sensitive to 
the proton structure, therefore, the shape coexistence of the neutron
structure may depend on the proton number. 
If a oblate neutron shape is favored also in neutron-rich nucleus
as well as in $^{28}$Si, one may meet another question: is 
the proton shape consistent with the neutron one ?
In this sense, $^{24}$Ne is an attractive nucleus, where the 
proton shape would be different from the neutron shape,  
because the prolate shape of proton density is favored in $Z=10$ systems
as seen in $^{20}$Ne.

In the study of proton and neutron deformations, the experimental
data of electric moments such as $B(E2)$ are useful to extract 
information about the intrinsic deformations. 
Recently, $B(E2;2^+\rightarrow 0^+)$ in $^{24}$Si has been measured by 
Coulomb excitation \cite{Kanno}. 
The measured $B(E2)$ in $^{24}$Si
is almost as large as the experimental value of $B(E2)$ in $^{24}$Ne. 
The ratio $B(E2;^{24}$Si)/$B(E2;^{24}$Ne)$\le 1$ is 
much smaller than a naive expectation based on a collective 
model picture that $B(E2)$ is proportional to $(N/Z)^2\approx 2$.  
If these nuclei are mirror symmetric, the ratio 
$B(E2;^{24}$Si)/$B(E2;^{24}$Ne)$\le 1$ lead to a possible difference 
between proton and neutron deformations in $^{24}$Ne.

The coexistence of the oblate and prolate
solutions in $^{28}$Si has been confirmed by many theoretical calculations
such as Nilsson-Strutinsky calculations\cite{Leander75},
Hartree-Fock-Bogoliubov\cite{Goodman}, 
alpha-cluster model approaches\cite{Bauhoff82a,Zhang94}.
Although mean field approaches are useful for 
systematic study of deformations, their applicability to 
very light nuclei is not obvious.
In the light nuclear region, we would like to remind the readers 
of the importance of cluster aspect, which closely relates
with the deformations.
The cluster aspect has been suggested also in light neutron-rich nuclei
as well as stable nuclei.
For example, 
it is well known that $^{20}$Ne has a $^{16}$O+$\alpha$ cluster structure, 
while
the development of clustering was suggested in neutron-rich 
B isotopes. 
The cluster aspect and the shape coexistence of even-even $N=Z$ nuclei
has been studied by $\alpha$-cluster models \cite{Zhang94,Bauhoff82a}. 
7$\alpha$-cluster model calculations were applied to $^{28}$Si, and
succeeded to describe an exotic shape with $D_{5h}$ symmetry
\cite{Bauhoff82a,Bauhoff82}, 
which is associated with the $K^\pi=5^-$ band observed in 
gamma transitions\cite{GLATZ}. Although the properties of the
oblate and prolate states were described by the 
7$\alpha$-cluster models \cite{Zhang94,Bauhoff82a}, however, 
many 7$\alpha$-cluster calculations failed to 
reproduce the order of the 
oblate and prolate solutions in $^{28}$Si. Namely, in most of the 
$\alpha$-cluster calculations, the prolate solutions are lower than the
oblate solutions
except for few calculations with Brink-Beoker forces \cite{Bauhoff82a}.
For the oblate property of the ground state in $^{28}$Si, it 
is important to incorporate the effect of spin-orbit force, which
is omitted within the $\alpha$-cluster models.
Furthermore, the cluster models are not suitable for 
the systematic study of unstable nuclei, because they
replies on the assumption of existence of cluster cores. 

For the systematic structure study of light nuclei,
one of the powerful approaches is the method of 
antisymmetrized molecular dynamics
(AMD). 
The applicability of this method for unstable 
nuclei has been proved in many
works \cite{ENYObc,ENYOsup,AMDrev,Thiamova03}.
In addition to description of deformations and cluster aspect
in light nuclei, this method has an advantage that electric moments
can be directly related with structure change 
based on the microscopic treatment of spin-parity projection  
without introducing effective charges.

In this paper, we study the deformations of $N=14$ isotones while 
focusing on the coexistence of the oblate and prolate neutron structure.
We pay attention to reproduction of the order of oblate and
prolate solutions, which coexist in $^{28}$Si, and discuss 
the effect of the spin-orbit force.
We analyze the systematics of deformations in the neutron excess nuclei 
in relation to the observables such as $Q$-moments and $E2$ transitions 
strength.
This paper is organized as follows. In the next section,
the formulation of AMD is 
briefly explained. We show the theoretical results and 
give comparisons with the experimental data in Sec.\ref{sec:results}. 
In Sec.\ref{sec:discuss}, we analyze the instrinsic structure, 
and discuss the lowering mechanism of the oblate state 
within the AMD framework.
Finally, a summary is given in Sec.\ref{sec:summary}.

\section{Formulation}
 \label{sec:formulation}

Here we briefly explain the formulations of the present calculations.
Details of the formulation of AMD methods for nuclear structure 
studies are explained in Refs.\cite{ENYObc,AMDrev}.
The present calculations are basically same as those
in Ref.\cite{ENYObc}.

The wave function of a system with a mass number $A$ 
is written by a superposition of AMD
wave functions $\Phi_{\rm AMD}$. An AMD wave function is given by a single
Slator determinant of Gaussian wave packets as,
\begin{equation}
 \Phi_{\rm AMD}({\bf Z}) = \frac{1}{\sqrt{A!}} {\cal{A}} \{
  \varphi_1,\varphi_2,...,\varphi_A \},
\end{equation}
where the $i$-th single-particle wave function is written as,
\begin{eqnarray}
 \varphi_i&=& \phi_{{\bf X}_i}\chi_i\tau_i,\\
 \phi_{{\bf X}_i}({\bf r}_j) &\propto& 
\exp\bigl\{-\nu({\bf r}_j-\frac{{\bf X}_i}{\sqrt{\nu}})^2\bigr\},
\label{eq:spatial}\\
 \chi_i &=& (\frac{1}{2}+\xi_i)\chi_{\uparrow}
 + (\frac{1}{2}-\xi_i)\chi_{\downarrow}.
\end{eqnarray}
The iso-spin function $\tau_i$ is fixed to be up(proton) or down(neutron),
and the  orientation of intrinsic spin 
$\xi_i$ is fixed to be $1/2$ or $-1/2$ in the present calculations
as done in Ref.\cite{ENYObc}.
The spatial part is represented by 
complex variational parameters, ${\rm X}_{1i}$, ${\rm X}_{2i}$, 
${\rm X}_{3i}$, which indicate the centers of Gaussian wave packets.

In the AMD model, all the centers of single-nucleon Gaussians 
are treated independently as the complex variational parameters.
Thus, this method is based completely on single nucleons and therefore
it does not rely on the assumption of the existence of cluster cores.
In the sense that a single AMD wave function is written by 
a Slator determinant of Gaussians, 
the AMD method is regarded as an extended model of 
Bloch-Brink cluster model\cite{brink66}. 
Here we note that the Gaussian center is expressed by 
the 'complex' parameter ${\bf X}_i$ which contains a real part 
and imaginary part. It means an extension of the model space, 
because the degrees of freedom are twice of the case that 
Gaussian centers are given by real values as in usual cluster models.
If we ignore the effect of antisymmetrization, 
the position ${\bf D}_i$ and the momentum ${\bf K}_i$ of the 
$i$-th single-nucleon wave packets are expressed 
by the real and imaginary parts of ${\bf X}_i$, respectively as,
\begin{eqnarray}
{\bf D}_i\equiv \frac{\langle \phi_{{\bf X}_i}|{\bf r}|\phi_{{\bf X}_i}\rangle}
{\langle \phi_{{\bf X}_i}|\phi_{{\bf X}_i}\rangle} &=&
\frac{{\rm Re}({\bf X}_i)}{\sqrt{\nu}} \nonumber\\
{\bf K}_i\equiv 
\frac{\langle \phi_{{\bf X}_i}|{\bf p}|\phi_{{\bf X}_i}\rangle}
{\langle \phi_{{\bf X}_i}|\phi_{{\bf X}_i}\rangle} &=&
2\hbar\sqrt{\nu}{\rm Im}({\bf X}_i).\label{eq:DK}
\end{eqnarray}
In the nuclear structure study with the AMD,
the imaginary parts of {\bf Z}
are essential to describe the rotation motion of the system. 
They are important to
incorporate the effect of spin-orbit force and to describe high-spin states.

We perform energy variation for a parity-eigen state, 
$P^\pm\Phi_{\rm AMD}\equiv \Phi^\pm_{\rm AMD}$, 
projected from an AMD wave function
by using the frictional cooling method\cite{ENYObc}.
We consider the AMD wave function obtained by the energy variation 
as the intrinsic state, and 
total-angular-momentum projection($P^J_{MK}$) is 
performed after the variation to calculate the expectation values of 
operators such as energies and moments. 
In the present calculations, 
the parity projection is done before the variation, but the 
total-angular-momentum projection is performed after the variation. 
In many of $N=14$ isotones, two local minimum solutions are found in the 
energy variation. 
In such cases, we diagonalize the Hamiltonian and norm matrices,
$\langle P^J_{MK'}\Phi^{'\pm}_{\rm AMD}|H| 
P^J_{MK''}\Phi^{''\pm}_{\rm AMD}\rangle$
and $\langle P^J_{MK'}\Phi^{'\pm}_{\rm AMD}| 
P^J_{MK''}\Phi^{''\pm}_{\rm AMD}\rangle$
with respect to the obtained intrinsic wave functions 
($\Phi^{'}_{\rm AMD}$,$\Phi^{''}_{\rm AMD}$) and 
the $K$-quantum ($K',K''$). After the diagonalization the 
ground and excited bands are obtained.
 
\section{Interactions} 
\label{sec:interaction}

The effective nuclear interactions adopted in the present work
consist of the central force, the
 spin-orbit force and Coulomb force.
We adopt MV1 force \cite{TOHSAKI} as the central force.
The MV1 force contains a zero-range three-body force 
in addition to the two-body interaction.
The Bertlett and Heisenberg terms are chosen to be $b=h=0$.
We use a parameter set, case 1 of MV1 force with the Majorana 
parameter $m=0.62$.
Concerning the spin-orbit force, 
the same form of the two-range Gaussian 
as the G3RS force \cite{LS}
is adopted. The strengths of the spin-orbit force, 
$u_{I}=-u_{II}\equiv u_{ls}=900$, and $u_{ls}=2800$ MeV are used. 

\section{Results}\label{sec:results}

We calculate the natural parity states of 
$N=14$ nuclei,$^{19}$B, $^{20}$C, $^{22}$O, $^{24}$Ne and $^{28}$Si,
with the AMD method. 
The width parameters
$\nu$=0.145, 0.15, 0.150, 0.155, 0.15 are chosen for 
$^{19}$B, $^{20}$C, $^{22}$O, $^{24}$Ne and $^{28}$Si,
respectively, so as
to minimize the energy of each nucleus.  

In these $N=14$ nuclei, 
we find two local minimum solutions with oblate and prolate
deformations of neutron density,
except for the results of $^{20}$C and $^{22}$O
with $u_{ls}=2800$ MeV.
It signifies the trend of the shape coexistence of neutron structure 
in the $N=14$ isotones.
The feature of the shape coexistence originates in 
the nature of $N=14$ neutron structure.

After performing the total-angular-momentum projection 
and the diagonalization for the obtained oblate and prolate states,
we obtain rotational bands, $K^\pi=3/2^-_1$ and $3/2^-_2$ bands 
in $^{19}$B, and $K^\pi=0^+_1$ and $0^+_2$ bands in $^{24}$Ne and $^{28}$Si
from the two local minimum solutions.
In the calculations of $^{20}$C and $^{22}$O with the stronger spin-orbit 
force $u_{ls}=2800$ MeV,
the shape coexistence phenomena disappear and 
only one intrinsic state is obtained 
by the energy variation in each nucleus.
In the calculations with $u_{ls}=2800$ MeV,
the lowest states of $^{22}$O and $^{24}$Ne have small prolate deformations 
of neutron density. 
As mentioned later,
these states with 'small prolate deformations' should be classified as
the 'oblate' states in the analysis of the shape coexistence in
$N=14$ systems, because their properties 
are similar to those of the oblate states of 
other $N=14$ nuclei.
 Therefore, we call these lowest states of $^{22}$O
and $^{24}$Ne as 'oblate' states in this paper.

\begin{figure}
\noindent
\epsfxsize=0.4\textwidth
\centerline{\epsffile{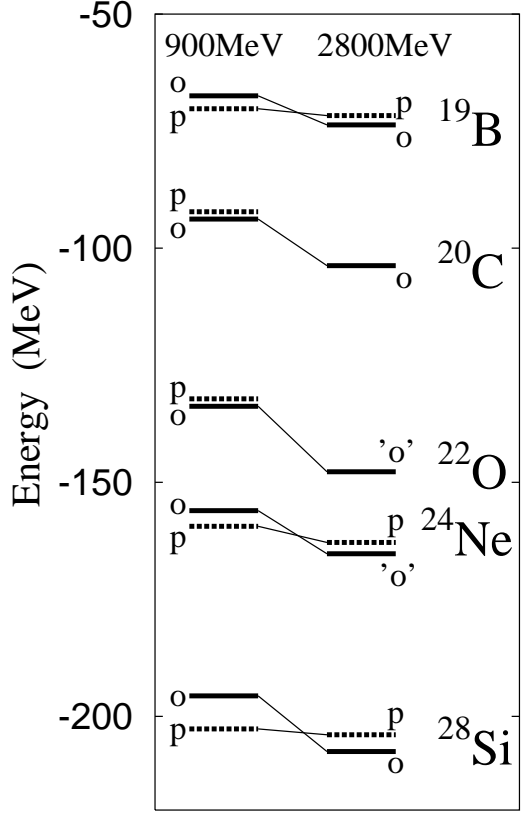}}
\caption{\label{fig:n14ls}
Calculated binding energies for the oblate(o) and prolate(p) 
states. The energies are of the band-head states, 
$J^\pm=3/2^-$ in $^{19}$B and the $0^+$ states 
in $^{20}$C, $^{22}$O, $^{24}$Ne, and $^{28}$Si.
The results with $u_{ls}=900$ MeV($u_{ls}=2800$ MeV) are shown in
left(right).  
Although the 'oblate'('o') states in $^{22}$O and $^{24}$Ne
have small prolate deformations of neutron density, these states are
attributed to the oblate neutron structure in $N=14$ systems.
The details are described in the text.}
\end{figure}

The energies of the band-head states are shown in Fig.\ref{fig:n14ls}. 
In the results with $u_{ls}=900$ MeV, we find that oblate and prolate
bands coexist in those $N=14$ nuclei. The oblate and prolate bands 
are almost degenerate in $^{20}$C and $^{22}$O, while in 
$^{19}$B, $^{24}$Ne and $^{28}$Si the prolate bands are lower than the 
oblate ones.
With the stronger spin-orbit force $u_{ls}=2800$ MeV, the 'oblate' solutions
are the lowest in all of these nuclei. Namely, 
the ground bands have the 'oblate' neutron shapes .
The prolate bands appear as excited bands
in $^{19}$B, $^{24}$Ne and $^{28}$Si.
We can not find prolate solutions in $^{20}$C and $^{22}$O
in the calculation with $u_{ls}=2800$ MeV.
 
The ground state of $^{28}$Si is experimentally known 
to be oblate. The order of the oblate and prolate bands
is well reproduced by the calculations with $u_{ls}=2800$ MeV.
On the other hand, the calculations with $u_{ls}=900$ MeV 
tend to relatively overestimate the energy of the oblate bands, and they  
fail to reproduce the order of the oblate and prolate solutions in $^{28}$Si.
Comparing the results with $u_{ls}=900$ MeV and $u_{ls}=2800$ MeV, 
it is found that the relative energy of 
the oblate and prolate states in $N=14$ system
is very sensitive to the strength of the spin-orbit force.
In order to describe the oblate feature of the ground state 
of $^{28}$Si, the effect of the spin-orbit force is 
significant as follows.
In comparison of the results for the stronger 
spin-orbit force($u_{ls}=2800$ MeV) with those for 
weaker one($u_{ls}=900$ MeV) in Fig.\ref{fig:n14ls}, we find that
the energy gains in the oblate states are as large as about 10 MeV,
while the energies of prolate states are almost unchanged
with the increase of the strength of the spin-orbit force.
As shown in table \ref{tab:ls}, the absolute value of the spin-orbit term
is much larger in the oblate states than in the prolate states.
It indicates that the energy gain of the spin-orbit force in the 
oblate state is essential to describe the level structure of $^{28}$Si.
The stronger spin-orbit force $u_{ls}=2800$ 
is appropriate in the present framework
for describing the order of the oblate and prolate levels in $^{28}$Si.
We discuss the lowering mechanism of the oblate states 
in section \ref{sec:discuss}.

\begin{figure}
\noindent
\epsfxsize=0.4\textwidth
\centerline{\epsffile{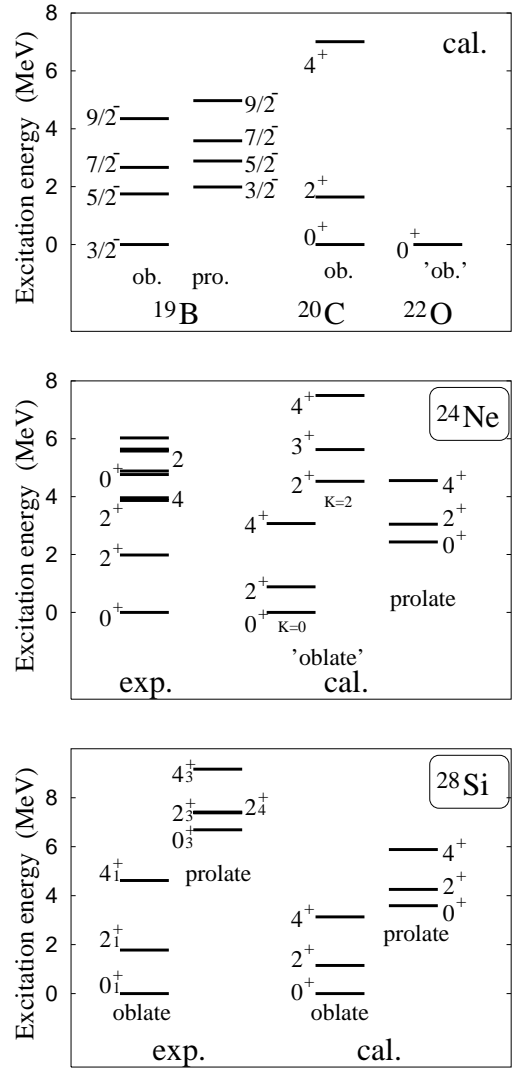}}
\caption{\label{fig:n14spe}
The level structure of $^{19}$B, $^{20}$C, $^{22}$O, $^{24}$Ne and $^{28}$Si
calculated with $u_{ls}$=2800 MeV. 
}
\end{figure}

Next, we show the level scheme obtained with $u_{ls}=2800$ MeV 
in Fig.\ref{fig:n14spe}.

In $^{19}$B, we predict the coexistence of oblate and prolate bands.
In the previous AMD calculations of B isotopes\cite{ENYObc}, 
the prolate shape in the ground state 
was predicted in $^{19}$B. This is because the adopted spin-orbit force 
was too weak in the previous calculations. On the other hand, when we use the
stronger spin-orbit force, which reproduces the order of the oblate 
and prolate bands in $^{28}$Si, 
we obtain the ground state of $^{19}$B with the oblate neutron shape 
while the prolately deformed state appears above the ground state.
As shown later, the excited prolate band has $^{11}$Li+$^8$He-like 
cluster structure. 
Although the 2-neutron separation energy of $^{19}$B 
is very small as about 0.1 MeV, 
we expect that the prolate states may exist as resonances
because such cluster states might be stable against neutron decays.

In $^{22}$O, we obtain the ground state but can not find 
other excited states in the present calculations.
This is because of the double shell-closed feature of $^{22}$O.
Namely, the intrinsic state has 
mostly spherical shape among the $N=14$ nuclei
due to the effects of the neutron $d_{5/2}$ sub-shell closure 
and the proton magic number $Z=8$.

In the results of $^{24}$Ne, 
the lowest 'oblate' band in the theoretical results corresponds to 
the ground band of experimental data as discussed in the next section.
On the other hand, the excited prolate bands are 
not experimentally identified yet.
In the present calculations, a side-band $K^\pm=2^+$ of the oblate 
ground is suggested in addition to the excited prolate band.

In $^{28}$Si, the shape coexistence of the oblate ground band and the prolate
excited band has been known.
The prolate band, which starts from the band-head $0^+_3$ state
at 6.691 MeV, has been identified in gamma-transition 
measurements\cite{GLATZ}. The $0^+_2$ state at 4.979 MeV in the experimental
data is considered to be a vibrational excitation in the oblate state. 
The level structure of the oblate and prolate bands 
are reproduced by the present calculations.
The level spacing between oblate and prolate bands can be
reproduced in good agreement by using a slightly 
stronger spin-orbit force $u_{ls}=3200$ MeV which gives 7.3 MeV excitation
energy for the band-head state of the prolate band.

\begin{figure}
\noindent
\epsfxsize=0.49\textwidth
\centerline{\epsffile{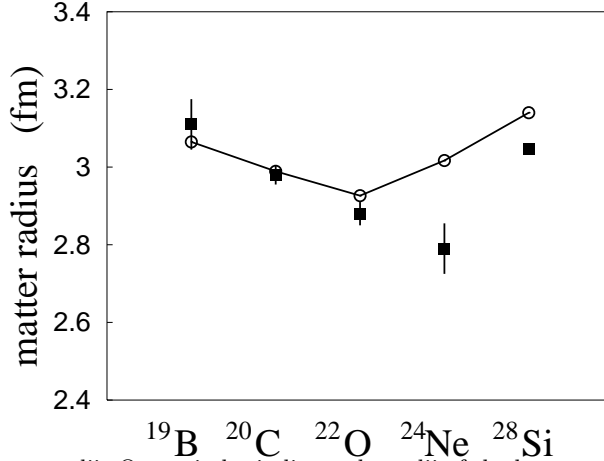}}
\caption{\label{fig:rmsr}
Root-mean-square matter radii.
Open circles indicate the radii of 
the lowest states calculated with 
$u_{ls}=2800$ MeV. 
The experimental data are shown by square points.
The data for $^{19}$B, $^{20}$C, $^{22}$O and $^{24}$Ne are those 
derived from the interaction 
cross sections \protect\cite{Ozawa}, and the radius of $^{28}$Si is 
derived from the charge radius taken from Ref.\protect\cite{VRIES} measured by 
electron scattering.
}
\end{figure}

The theoretical results of the 
root-mean-square matter radii, $E2$ transition strengths, and
electric and magnetic moments are shown in Fig.\ref{fig:rmsr} and 
tables \ref{tab:be2} and \ref{tab:qmom} in comparison with the experimental
data.  The present calculations reasonably agree to the experimental data.

\begin{table}
\caption{ \label{tab:ls} Expectation values(MeV) of the
spin-orbit force in the oblate and prolate states obtained with 
$u_{ls}=2800$ MeV. 
Each value is for the band-head state obtained by spin-parity
projection from an single AMD wave function
which corresponds to 'oblate' or prolate solution.
The superposition of the oblate and prolate solutions 
is not done.
}
\begin{center}
\begin{tabular}{cccccc}
 & $^{19}$B & $^{20}$C & $^{22}$O & $^{24}$Ne & $^{28}$Si \\
'oblate' & $-15.9$ & $-18.8$ & $-22.2$ & $-17.9$ & $-23.9$ \\ 
'prolate' & $-2.6$ & $-$ & $-$ & $-9.9$ & $-2.0$ \\
\end{tabular}
\end{center}
\end{table}

\begin{table}
\caption{ \label{tab:be2} $E2$ transition 
strengths.
The results are calculated with $u_{ls}$=2800 MeV.
The theoretical $B(E2)$ in $^{24}$Si are evaluated 
by assuming mirror symmetry with $^{24}$Ne.
The experimental data are taken from Refs.\protect\cite{GLATZ,Kanno,Raman}.
$^a$ Since the collective $4^+\rightarrow 2^+$ transition
in the $K^\pi=0^+_2$ band of $^{28}$Si is distributed over two transitions
into $2^+(7.381)$ and $2^+(7.417)$, the experimental value in this table is
a sum of the strengths of those two transitions in Ref.\protect\cite{GLATZ}.
}
\begin{center}
\begin{tabular}{cclc|clc}
 & & &  cal. &     &  &    exp.\\ 
   & band & transitions& $B(E2)$(w.u.) & band & $E_i\rightarrow E_f$ 
& $B(E2)$ (w.u.) \\
\hline
$^{28}$Si & oblate &$B(E2;2^+\rightarrow 0^+)$ & 9.1  &
 $K^\pi=0^+_1$  & 1.779$\rightarrow$0 & 12.9$\pm$0.5\\
          &        &$B(E2;4^+\rightarrow 2^+)$ & 12.8 &
 $K^\pi=0^+_1$  &4.617$\rightarrow$1.779  &13.6(+1.4,$-1.2$) \\
          & prolate&$B(E2;2^+\rightarrow 0^+)$ & 28.1 & 
          &        &                            \\
          &        &$B(E2;4^+\rightarrow 2^+)$ & 40.0 &
 $K^\pi=0^+_2$ & 9.164$\rightarrow$(7.381,7.417) &  43.5(+11.6,$-8.7$)$^a$ \\
\hline
$^{24}$Ne & 'oblate' &$B(E2;2^+\rightarrow 0^+)$& 9.2 &
 $K^\pi=0^+_1$  & 1.981$\rightarrow$0 & 6.8$\pm$ 2.9\\
    & prolate &$B(E2;2^+\rightarrow 0^+)$& 7.8 &
    &   &   \\
\hline
$^{24}$Si & 'oblate' &$B(E2;2^+\rightarrow 0^+)$& 9.4 &
 $K^\pi=0^+_1$  & 1.88$\rightarrow$0 & 4.6$\pm$ 1.4\\
    & prolate &$B(E2;2^+\rightarrow 0^+)$& 17 &
    &   &   \\
\hline
$^{20}$ C     & oblate &$B(E2;2^+\rightarrow 0^+)$& 1.8 &
    &   &   \\
\end{tabular}
\end{center}
\end{table}

\begin{table}
\caption{ \label{tab:qmom} Electric quadrupole moments and magnetic dipole
moments.
The theoretical results are calculated 
with $u_{ls}$=2800 MeV.
}
\begin{center}
\begin{tabular}{cccccc}
 &  & cal. & exp. \\
\hline
$^{28}$Si & $Q(2^+_1)$ (e$^2$fm$^4$) & 132 & 160(3) \\ 
 & $\mu(2^+_1)$ $(\mu_N)$ &  1.03 & 1.12(18) \\
\hline
 $^{19}$B & $Q(3/2^-_1$;oblate) (e$^2$fm$^4$)  & 34 &  \\ 
 & $Q(3/2^-_2$;prolate) (e$^2$fm$^4$)  & 43 &  \\ 
 & $\mu(3/2^-_1$;oblate) $(\mu_N)$  & 2.37 & \\ 
 & $\mu(3/2^-_2$;prolate) $(\mu_N)$  & 2.46 & \\ 
\end{tabular}
\end{center}
\end{table}

\section{Discussion}\label{sec:discuss}

In this section, we analyze the deformations 
of proton and neutron densities in the intrinsic states 
and discuss their effects on the observables such as $E2$ transitions.
The following discussions are based on the calculations with $u_{ls}$=2800 MeV.

\subsection{Intrinsic deformations}\label{subsec:intrinsic}

\begin{figure}
\noindent
\epsfxsize=0.35\textwidth
\centerline{\epsffile{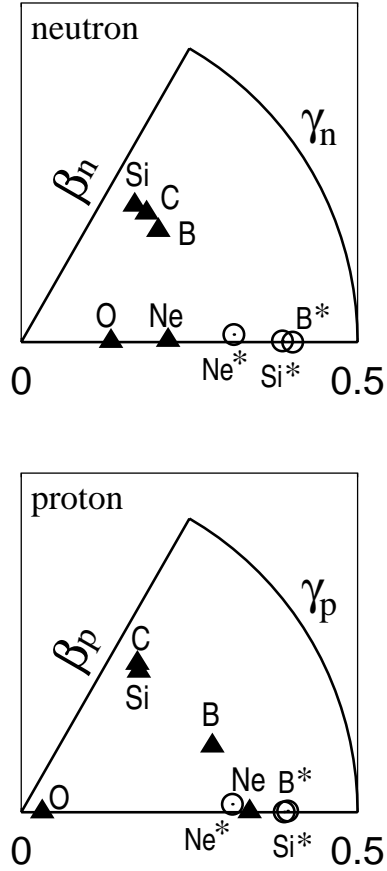}}
\caption{\label{fig:defo-n14}
Deformation parameters of the intrinsic states
$\Phi_{\rm AMD}({\bf Z})$ obtained with $u_{ls}=2800$ MeV. 
The deformation parameters $\beta_n,\gamma_n$($\beta_p,\gamma_p$) 
for the neutron(proton) 
densities are plotted in the upper(lower) panel. 
Triangles denote the deformations in the ground bands of 
$^{19}$B, $^{20}$C, $^{22}$O, $^{24}$Ne and $^{28}$Si,
while circles indicate those in the excited bands of 
$^{19}$B$^*$, $^{24}$Ne$^*$ and $^{28}$Si$^*$.}
\end{figure}

\begin{figure}
\noindent
\epsfxsize=0.35\textwidth
\centerline{\epsffile{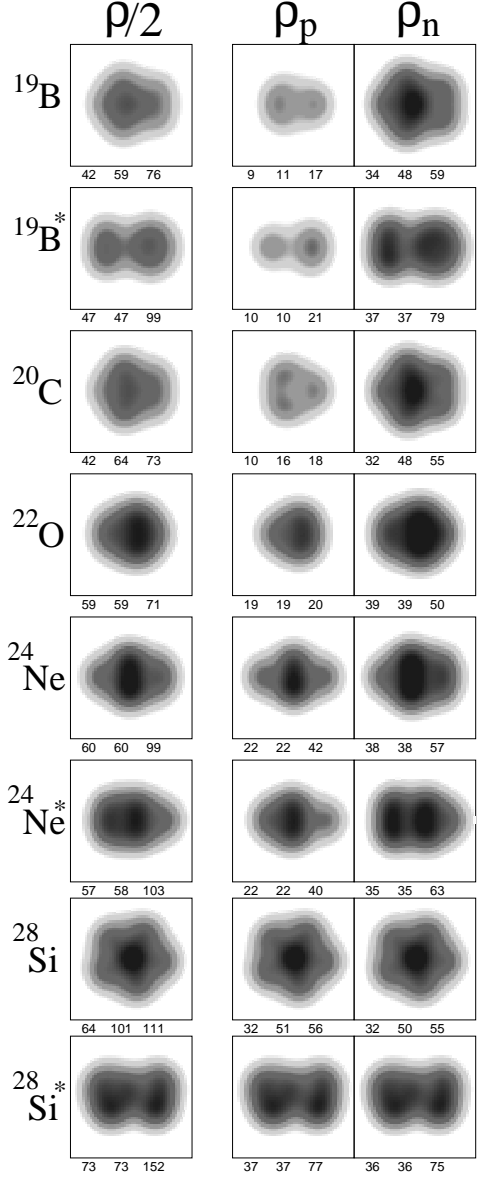}}
\caption{\label{fig:dense}
Density distribution of the intrinsic state 
$\Phi_{\rm AMD}({\bf Z})$. The axis of the intrinsic frame are chosen as
$\langle xx \rangle \le\langle yy \rangle 
\le\langle zz \rangle$ and $\langle xy \rangle =\langle yz \rangle =
\langle zx \rangle=0$.
The intrinsic system is projected 
onto the $zy$-plane.
The density is integrated along the transverse axis $x$.
The densities for matter, protons and neutrons are 
displayed in the left, middle 
and right figures, respectively. 
$\langle xx \rangle$, $\langle yy \rangle$, 
$\langle zz \rangle$
for matter, proton and neutron densities are written below 
the figures. The unit of the box frame size is 10 fm$\times$10 fm. 
}
\end{figure}

As mentioned before, two local minimum states 
are obtained in each system except for $^{20}$C and $^{22}$O.
The deformation parameters $\beta$, $\gamma$ for neutron and proton 
densities in the intrinsic states are shown in Fig.\ref{fig:defo-n14}.
The definition of $\beta$, $\gamma$ are given in Ref.\cite{ENYO-c10}.
Figure \ref{fig:dense} shows the distribution of matter, proton 
and neutron densities.
As seen in Figs.\ref{fig:defo-n14} and \ref{fig:dense}, 
the lowest states of $^{19}$B, $^{20}$C and $^{28}$Si have oblate neutron
shapes. Although the lowest states of $^{22}$O and $^{24}$Ne 
have small prolate deformations of neutron densities, we notice that
the characteristics of the neutron structure in these states are
rather similar to those in the oblate states of other $N=14$ nuclei. 
One of the remarkable features of the 'oblate' neutron structure is the 
larger energy gain of the spin-orbit force than the prolate neutron structure. 
In fact, the energy gains of the 
spin-orbit force in these states of $^{22}$O and $^{24}$Ne 
are as large as those in the 
oblate states in $^{19}$B, $^{20}$C and $^{28}$Si
as seen in Fig.\ref{fig:n14ls} and table \ref{tab:ls}.
It is considered that the oblately deformed neutron structure is not rigid 
but is somehow soft to vary into the small prolate deformation
in such systems as $^{22}$O and $^{24}$Ne, which have 
the spherical and prolate proton shapes, respectively.
In other words, the oblate neutron deformations can be modified 
because of the inconsistency with the proton deformations.
It should be stressed that the features of the shape coexistence
of neutron structure in $N=14$ nuclei varies depending on the proton number.

In $^{19}$B, the lowest state has the oblate neutron shape and 
a triaxial proton shape, while in the excited state
the large prolate deformation with
a cluster structure is enhanced.
As seen in Fig.\ref{fig:dense},
the developed $^{8}$He+$^{11}$Li-like clustering is suggested in 
the prolate excited band.

In the lowest state of $^{20}$C, both the proton and neutron shapes are
oblate. The reason for the absence of the prolate solution in $^{20}$C 
is understood as follows. 
As known in $^{12}$C, the proton shape is oblate in $Z=6$ nuclei.
Because of the inconsistency with the oblate proton deformation,
the prolate neutron structure is energetically unfavored 
comparing with the oblate neutron structure.

In $^{22}$O, the neutron deformation is the smallest among these $N=14$
isotones due to the
spherical proton shape which originates in the $p$-shell closure effect.
Another local minimum solution with the large prolate deformation of 
neutron structure 
does not appear as well as in $^{20}$C.

In $^{24}$Ne, two local minimum solutions with different neutron structures 
are obtained. The neutron structure of the lowest state has 
a smaller prolate deformation than that of the excited state.
This smaller neutron deformation in the lowest state
can be attributed to the oblate neutron structure in
$N=14$ systems, because as mentioned above
the sensitivity of the energy of this state to 
the strength of the spin-orbit force 
is quite similar to that of the oblate states in other $N=14$ isotones.
On the other hand, in the excited state, the neutron structure 
has a large prolate deformation and is similar to 
the prolate excited state of $^{28}$Si. 
Therefore, the coexistence of two local minimum solutions in $^{24}$Ne 
is associated with the shape coexistence of the oblate and prolate states 
in $^{28}$Si.
One of the unique features in $^{24}$Ne is that the 
proton deformations are prolate in both the ground and excited states.
It is because of the nature of $Z=10$ systems.
Due to the effect of prolate proton deformation, 
the oblate neutron structure varies into the small prolate deformation.
As a result, the deformation parameters $\beta_p$ and $\beta_n$ for the
proton and neutron densities are inconsistent with each other
as $\beta_p > \beta_n$.
This leads to a difference between proton and 
neutron deformations in the ground state of $^{24}$Ne.
The details of the different deformations
are discussed later in relation to the $E2$ transition 
strengths.

\begin{figure}
\noindent
\epsfxsize=0.3\textwidth
\centerline{\epsffile{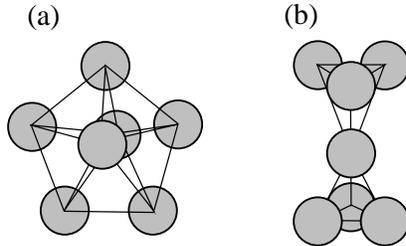}}
\caption{
Schematic figures for the positions(${\bf D}_i$) of the Gaussian 
centers in the oblate(a) and prolate(b) states of $^{28}$Si. 
The positions ${\bf D}_i(i=1\sim 28)$
are distributed in 7 groups which associate with the 7$\alpha$-cluster 
configurations. \label{fig:7alpha}
}
\end{figure}

In the results of $^{28}$Si, the oblate and prolate states 
coexist. As already mentioned,
the oblate feature of the ground state is reproduced with the stronger
spin-orbit force($u_{ls}=2800$ MeV). 
The shape coexistence of $^{28}$Si has been studied by 
7$\alpha$-cluster models \cite{Bauhoff82a,Zhang94,Bauhoff82}, where
exotic shapes have been suggested in addition to the normal oblate and
prolate deformations. 
For example, a pentagon shape composed of 7 $\alpha$ clusters with 
$D_{5h}$ symmetry \cite{Bauhoff82} is attributed to 
the $K^\pi=5^-$ rotational band which has been observed 
in gamma transition measurements \cite{GLATZ}.
In the present calculations, 
all the centers of 28 single-nucleon Gaussians are treated 
as independent variational parameters without assuming existence of 
any cluster cores.
By analysing the positions ${\bf D}_i(i=1\sim 28)$
of Gaussian centers, we find that the oblate and prolate states 
consist of 7 $\alpha$-like clusters in the present results.
Schematic figures for spatial
configurations of the oblate and prolate states are shown in
Fig.\ref{fig:7alpha}. Due to the 7$\alpha$ clustering, 
the pentagon shape appear
in the oblate solution as seen in Fig.\ref{fig:dense}. 
We give a comment on '$\alpha$' clusters in the present results.
Since the imaginary parts of the single-nucleon Gaussian centers
are non-zero in the oblate solution,
the $\alpha$-like clusters 
are {\it dissociated} $\alpha$ clusters which 
differ from the ideal $\alpha$ clusters
written by $(0s)^4$ harmonic-oscillator wave functions.
The details are discussed later in the last part of this section.

\subsection{$^{19}$B}

The structure of B isotopes has been studied with 
the AMD method in Ref.\cite{ENYObc}. 
In the previous results, it was predicted that cluster structure develops 
as a neutron number increases from $^{13}$B toward 
the neutron-drip line $^{19}$B
due to a large prolate deformation of neutron structure.
In the present results, however, the ground state of $^{19}$B has 
the oblate neutron structure, which is inconsistent with the previous results.
In the previous calculations of $^{19}$B, 
the oblate solution was not lower than the 
prolate solution because the adopted spin-orbit force $u_{ls}$=900 MeV 
was too weak.
As explained before, the relative energy of the prolate and
oblate solutions is sensitive to the strength of spin-orbit force.
Therefore, it is important to use a proper strength of 
the spin-orbit force to describe the feature of 
shape coexistence in $N=14$ systems. By using the stronger spin-orbit force
$u_{ls}$=2800 MeV, which can reproduces the level scheme of the oblate 
and prolate bands in $^{28}$Si, the 
oblate band becomes lowest in $^{19}$B while the prolate band appear as the 
excited band (Fig.\ref{fig:n14spe}). 
As a result, the neutron structure of 
the ground state of $^{19}$B is predicted to be oblate in the present work.
The oblate property of the ground band of $^{19}$B is contrast to the features
of other neutron-rich B isotopes($^{15}$B and $^{17}$B), 
where prolate neutron structure is favored\cite{ENYObc}.

Because of the oblate neutron structure in $^{19}$B, 
the proton deformation is smaller, and the clustering is weaker in the
ground state than the excited prolate state as seen in Fig.\ref{fig:dense}.
The deformation of the neutron structure is reflected in the electric quadrupole
moment($Q$) through the proton deformation. As shown in table \ref{tab:qmom},
the calculated $Q(3/2^-$) for the oblate ground state
is much smaller than that for the prolate excited state.
The experimental measurement of the $Q$ moment of $^{19}$B is required to
confirm the oblate neutron deformation.  
Contrary to the $Q$ moments, the calculated 
magnetic dipole moments $\mu$ are not so much different between oblate 
and prolate states.

\subsection{$^{24}$Ne}
$^{24}$Ne is an interesting nucleus, 
because the proton structure in $Z=10$ system may favor a
prolate shape as known in $^{20}$Ne.
Considering the coexistence of oblate and prolate shapes 
in $N=14$ neutron structure, 
it is natural to expect that 
the neutron shape and the proton shape may compete.  
The present results suggest 
that the different kinds of neutron structure coexist 
in $^{24}$Ne. One is the smaller prolate deformation in the ground band,
and the other is the larger prolate deformation in the excited band. 
The former one is attributed to the oblate structure in the $N=14$ systems.
We consider that the change of the neutron structure 
from the oblate shape into the small prolate deformation is understood as
the oblate neutron structure is somehow soft and is varied by 
the prolate nature of the proton structure. 
The present results suggest the smaller neutron deformation $\beta_n$
than the proton deformation $\beta_p$ as $\beta_n<\beta_p$ 
in the ground band of $^{24}$Ne.
On the other hand, $\beta_n$ is as large as $\beta_p$ 
in the excited band. 

In order to extract information about the proton and neutron deformations 
from the experimental data,
it is useful to analyze $E2$ transition strengths and compare them 
with those of mirror nuclei. 
Similar analyses on the electric moments were done
in Ref.\cite{ENYO-c10,ENYO-c16} where deformations of C isotopes have been 
discussed.
Recently, $B(E2;2^+_1\rightarrow 0^+_1)$ in $^{24}$Si has 
been measured by Coulomb excitation \cite{Kanno} as 
$B(E2)=4.6\pm 1.4$ w.u., which is smaller than or of same order of  
$B(E2;2^+_1\rightarrow 0^+_1)=$6.8$\pm$ 1.6 w.u. 
in $^{24}$Ne.
If we assume the mirror symmetry between $^{24}$Ne and $^{24}$Si,
$B(E2;^{24}$Si) in $^{24}$Si should equal to $B_n(E2;^{24}$Ne) 
for neutron transitions in $^{24}$Ne. Therefore,
these facts suggest that, in $^{24}$Ne, 
the neutron transition strength $B_n(E2;^{24}$Ne) is as small as 
the proton transition strength $B(E2;^{24}$Ne).
The ratio $B_n(E2;^{24}$Ne)$/B(E2;^{24}$Ne)$ \le 1$ is 
smaller than the naive expectation, $(N/Z)^2\approx 2$, 
given by a collective model picture.
The reduction of $B_n(E2;^{24}$Ne) can be described by the smaller neutron 
deformation $\beta_n$ than the proton deformation $\beta_p$ in the 
'oblate' ground band. In fact, the calculated ratio 
$B_n(E2)/B(E2)=B(E2;^{24}$Si)/$B(E2;^{24}$Ne) in the 'oblate' band 
of $^{24}$Ne is approximately 1 (table \ref{tab:be2}), which well agrees to
the experimental data.
 On the other hand, in the excited prolate band with 
$\beta_n\approx \beta_p$, the calculated $B(E2;^{24}$Si$^*$) is 
twice larger than the $B(E2;^{24}$Ne$^*$), and 
the ratio $B_n(E2)/B(E2)=B(E2;^{24}$Si$^*$)/$B(E2;^{24}$Ne$^*$) 
is consistent with the collective model expectation $(N/Z)^2\approx 2$.
Comparing the calculations and the experimental data of
$B(E2;^{24}$Ne) and $B(E2;^{24}$Si),
we conclude that the 'oblate' band correspond to the ground band of $^{24}$Ne.
The present result of the 'oblate' band 
still overestimates the $B(E2;^{24}$Si) as shown in table \ref{tab:be2}.
It is conjectured that the neutron deformation might be smaller due to the  
$d_{5/2}$-shell closure effect.

\subsection{Lowering mechanism of oblate states}

The shape coexistence in $^{28}$Si has been studied by 
7$\alpha$-cluster models for a long time\cite{Bauhoff82a,Zhang94}.
The oblate and prolate solutions coexist as local minima 
in the $\alpha$-cluster model space, however,  
most of the calculations fail to 
reproduce the order of the oblate and prolate states in $^{28}$Si.
Even in the three-dimensional calculations of 
7$\alpha$ clusters, the prolate solution is lower than 
the oblate solution \cite{Zhang94}, which is inconsistent with the
fact of the oblate ground state.
Exceptions are the 7$\alpha$ calculations with 
a few parameter sets of Brink-Beoker forces in Ref.\cite{Bauhoff82a}, 
though the binding energy is much underestimated.

The main reason for the failures in describing the deformation of the 
ground state of $^{28}$Si in the 7$\alpha$-cluster calculations
is because the spin-orbit force is
omitted in the $\alpha$-cluster models.
As seen in the Nilsson orbits\cite{Bohr75}, 
the spin-orbit force plays an important role to make 
the shell gap at $N=14$ in the oblate system.
It means that the spin-orbit force is significant 
to gain the energies of the oblate states in $N=14$ systems.
Here we explain the lowering mechanism of the oblate state in 
$^{28}$Si within the AMD model space.

\begin{figure}
\noindent
\epsfxsize=0.4\textwidth
\centerline{\epsffile{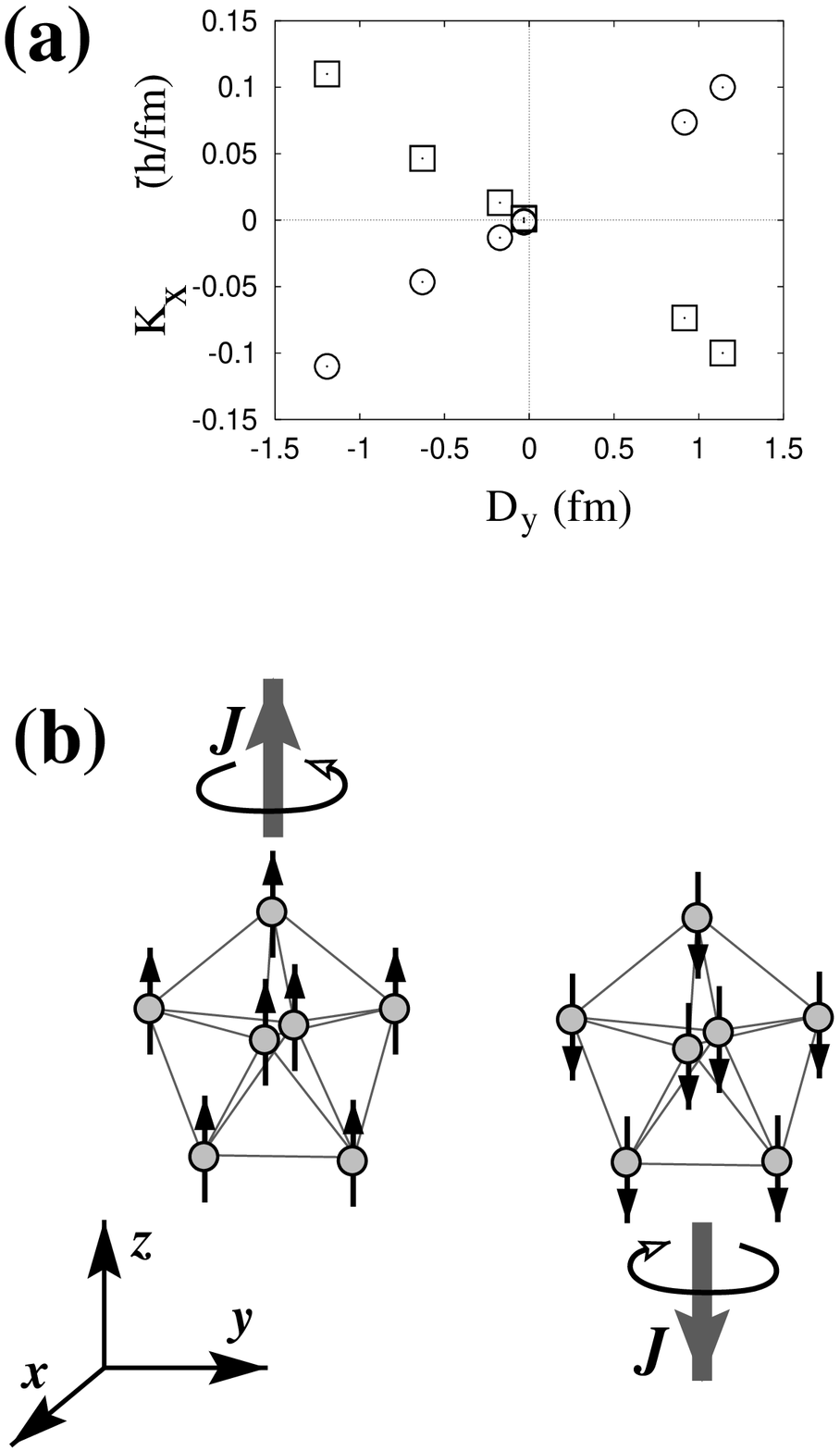}}
\caption{\label{fig:rotation}
(a) Positions $D_y$ and momenta $K_y$ of 28 Gaussian centers
in the intrinsic wave function of the oblate solution of $^{28}$Si.
The $x$, $y$ and $z$ axis are chosen to be $\langle x^2\rangle \le
\langle y^2\rangle\le\langle z^2\rangle$. It is found that 
the $z$-axis in the oblate solution 
almost equals to the direction of the intrinsic-spins of
single-nucleon wave functions.
(b) Schematic figure for the rotational motion given by
the positions and momenta of Gaussian centers.
}
\end{figure}

As described in \ref{subsec:intrinsic}, in the results
 of $^{28}$Si it is found that
the positions ${\bf D}_i$(the real part of ${\bf X}_i/\sqrt{\nu}$) 
of Gaussian centers correspond to the 7$\alpha$-like cluster
configurations (Fig.\ref{fig:7alpha}).
If the imaginary parts of the Gaussian centers are zero, the expectation value
of the spin-orbit force vanishes because it is exactly zero in the 
systems composed of the ideal $(0s)^4$-$\alpha$ clusters. 
However, in the intrinsic state of the oblate band, we find that
the non-zero imaginary parts of the Gaussian centers cause the rotational 
motion.
The positions and the momenta 
of the single-nucleon Gaussian wave packets are given by the real parts
$\{D_y\}_i$ and imaginary parts $\{K_x\}_i$ 
of the Gaussian centers (${\bf X}_i/\sqrt{\nu}$) as given 
in Eq.\ref{eq:DK}.
$\{D_y\}_i$ and $\{K_x\}_i$ in the intrinsic state of the oblate ground 
band of $^{28}$Si are plotted in Fig.\ref{fig:rotation}. 
The positions and momenta for up-spin and down-spin nucleons 
are shown by square and circle points, respectively. 
The correlation between 
$D_y$ and $K_x$ indicates the rotational motion around the $z$-axis. 
Namely, the angular momenta of the up-spin nucleons are parallel to the 
$z$-axis and 
those of the down-spin nucleons are anti-parallel. 
Therefore, the rotational motions
for up-spin nucleons and the down-spin nucleons are reverse to each other, 
and make the spin-orbit force attractive. The energy gain 
of the spin-orbit force is more than $20$ MeV as shown in table \ref{tab:ls}.
In the $k$-th $\alpha$-like cluster, 
the momenta $K_x(\uparrow)$ for the up-spin proton and neutron are 
opposite to those $K_x(\downarrow)$ for the down-spin proton and
neutron as $K_x(\uparrow)=-K_x(\downarrow)$
as shown in Fig.\ref{fig:rotation}. 
Although the positions ${\bf D}_i$ are located in the
$7\alpha$-like cluster configurations, it is important that the 
$\alpha$-like clusters are dissociated due to the imaginary parts ${\bf K}_i$,
because the Gaussians for the up-spin nucleons 
are moving in the reverse to the 
down-spin nucleons in each of $\alpha$-like clusters. 

Thus, in the AMD framework, the spin-orbit force can be gain 
while keeping the $\alpha$-like cluster configurations.
The energy gain of the spin-orbit force arises from the flexibility of the
AMD wave functions where Gaussian centers are expressed by
'complex' variational parameters instead of real parameters.

\section{summary}\label{sec:summary}
We studied the deformations of $N=14$ isotones, $^{19}$B, $^{20}$C,
$^{22}$O, $^{24}$Ne and $^{28}$Si, while focusing on the 
shape coexistence of the oblate and prolate neutron structure.
The relative energies between the oblate and prolate states are found to be
sensitive to the strength of spin-orbit force. 
By using a set of interaction parameters $m=0.62$ and $u_{ls}=2800$ MeV
in the MV1(case 1)+G3RS force, we can describe
the properties of the shape coexistence in $^{28}$Si, which is known to have
the oblate ground band and the prolate excited band.
The present results reasonably agree to the 
experimental data of radii, moments, and $E2$ transition strengths.

In the present calculations, the solutions with 
the oblate neutron structure are energetically favored and form the ground
bands in these $N=14$ isotones.
In the results of $^{19}$B, the prolately deformed excited
band with cluster structure was predicted to appear
above the oblate ground state. 
In $^{24}$Ne, we proposed a possible 
difference between proton and neutron deformations in the ground band.
We discussed $Q$-moments and $E2$ transition strengths 
in relation to the intrinsic deformations.
In the analysis of the results, it was found that the spin-orbit force plays
an important role in the lowering mechanism of the oblate neutron
structure in $N=14$ systems. We found that the imaginary parts of the
single-nucleon Gaussian centers in the AMD model space are important
to gain the spin-orbit force.

\acknowledgments

The author would like to thank Prof. H. Horiuchi, Prof. K. Ikeda, and
Dr. M. Kimura for many discussions. She is also thankful to 
Prof. T. Motobayashi for valuable comments. 
The computational calculations in this work were supported by the 
Supercomputer Project Nos. 70, 83 and 93
of High Energy Accelerator Research Organization(KEK).
This work was supported by Japan Society for the Promotion of 
Science and a Grant-in-Aid for Scientific Research of the Japan
Ministry of Education, Science and Culture.
The work was partially performed in the ``Research Project for Study of
Unstable Nuclei from Nuclear Cluster Aspects'' sponsored by
Institute of Physical and Chemical Research (RIKEN).

\end{document}